\documentclass[12pt,a4paper]{article}

 \usepackage[dvips]{graphics}
 \usepackage{graphicx}
 \usepackage{afterpage}
 \usepackage{enumerate}
\begin{document}

\title{\bf Absolute shifts of Fe~I and Fe~II lines\\ in solar active regions (disk center)
 }
 \author{\bf P.N. Brandt$^1$, A.S. Gadun$^2$, V.A. Sheminova$^2$ }
 \date{}

 \maketitle
 \thanks{}
\begin{center}
{$^1$Kiepenheuer-Institut  f\"{u}r  Sonnenphysik,
Sch\"{o}neckstr. 6, D-7800 Freiburg\\ Federal Republik of
Germany}\\ {$^2$Main Astronomical Observatory, National Academy
of Sciences of Ukraine
\\ Zabolotnoho 27, 03689 Kyiv, Ukraine}\\
\end{center}

 \begin{abstract}
We estimated absolute shifts of Fe~I and Fe~II lines 
from Fourier-transform spectra observed in solar active regions. 
Weak Fe~I lines
and all Fe~II lines tend to be red-shifted as compared to their
positions in quiet areas, while strong Fe~I lines, whose cores
are formed above the level $\log \tau_5\approx-3$   (about
425~km), are relatively blue-shifted, the shift growing with
decreasing lower excitation potential. We interpret the results
through two-dimensional MHD models, which adequately reproduce
red shifts of the lines formed deep in the photosphere. Blue
shifts of the lines formed in higher layer do not gain substance
from the models.
\end{abstract}

\section{Introduction}

We continue the study initiated by us in the papers \cite{4,5} 
concerned with the effect of small-scale magnetic fields 
on solar granulation.
The study is based on the observations made by P. Brandt with
a Fourier-transform spectrometer (FTS) at the McMath telescope
(Kitt Peak National Observatory, USA). The advantages of FTS
observations are their high quality and a possibility to observe
almost concurrently a large number of lines. Areas close to the
solar disk center were observed.

Here we investigate in great depth absolute shifts of Fe~I and
Fe~II lines in low-activity areas (plages) with different
integral magnetic fluxes. This is the first time that such an
investigation covers a large number of lines formed at various
levels in the solar photosphere so that effects of small-scale
magnetic fields on height distribution of photospheric velocity
field may be studied in detail.

The present state of the investigations on small-scale solar
magnetic fields is elucidated in review  \cite{20} by Solanki.
Determinations of absolute shifts of absorption lines in active
areas are not numerous, they are overviewed in  \cite{1}.

Paper  \cite{7} by Brandt and Solanki, similar to  \cite{4,5}
and the present paper, is also based on the FTS observations, but
only 32 lines were used there to study line parameter variations
and 19 lines to determine absolute shifts. Those 19 lines were
strong Fe~I lines only. The conclusion was made that the lines
in active areas had red shifts with respect to their positions
in the quiet photosphere. The shifts depend on magnetic field
strength (filling factor) and are, on the average, 0.22 pm
(about 120~m/s) for lower bisector sections. This result agrees
qualitatively with the findings of the studies by Cavallini et
al.  \cite{8}, Livingston  \cite{17}, Immerschitt and Schroter
\cite{15}.

At the same time Cavallini et al.  \cite{9} dealing with
absolute shifts of four lines formed at different photospheric
levels obtained a quite different result --- shifts of line
cores with respect to their positions in the quiet photosphere
depend on the region of line core formation: the weak Fe~I line
at $\lambda$  614.92~nm displays a red shift, while the strong
Ca~I line at  $\lambda$ 616.22~nm is blue-shifted. A similar
result was obtained by Keil et al.  \cite{16} for the strong
Fe~I $\lambda$ 543.45~nm line with a small lower excitation
potential (1.01~eV), the core of this line is formed in the
upper photospheric layers. Absolute shifts of the line were
found to depend on magnetic field polarity, but when the shifts
are averaged without regard for local field signs, the line core
has a blue shift of about 150~m/s.

Evidence of different behavior of velocity fields with height in
the photosphere in active and quiet areas is found also in the
analysis of spectral observations with high spatial resolution
made by Hanslmeier et al.  \cite{13,14}. If this is the case,
lines formed at different levels in the atmosphere should
display different absolute shifts.

Thus, the problem needs to be thoroughly investigated on the
basis of an extended sample of spectral lines. This is the prime
objective of the present study.

\section{Observations}

The observations used in this study were made in June 1984. The
spectral region adequate for reduction extended from  $\lambda$ 505~nm to
 $\lambda$ 665~nm. The theoretical resolution  $\lambda$/$\Delta \lambda$
was about 200~000. To assure a sufficient stability of results,
the signal integration time during observations was no less than
13.7 min (lifetime of one granule) and the entrance slit was
$5^{\prime\prime}\times25^{\prime\prime}$ (it covered
approximately 50 granules). The observations and their comparison to the Liege Atlas
data  \cite{10} were described in  \cite{4,7}. In all, 23
spectrograms acquired for areas close to the disk center were
used in the study. 

{\bf The magnetic field strength} in the areas was
estimated by the filling factor $\alpha$ --- the relative solar surface
area occupied by small-scale magnetic features. The filling
factor for the observations used by us was calculated by Solanki
 \cite{7} from 182 Fe~I lines. A simple two-component model was adopted
for active areas, it consisted of a ``nonmagnetic'' quiet
photosphere area of size $1- \alpha$ and an area of size 
$\alpha$ occupied by magnetic elements with a strength of 0.15~T,
relative contrast in the continuum was 1.4, and average line
weakening was 0.7. The method was described in detail in
\cite{7}.

{\bf Line selection.} Lines for the study were selected from two
lists. The first list was compiled in  \cite{4,5} especially for
studying Fe~I and Fe~II line asymmetries from the same FTS
observations. It comprised 281 lines of Fe~I and 30 lines of
Fe~II between 505~nm and 665~nm. The second list was compiled by
Dravins et al.  \cite{11,12} for studying the fine structure of
Fe~I and Fe~II lines in the spectra of the quiet Sun. We used
the laboratory wavelengths given in this list. As a result of
critical selection we got 189 lines of Fe~I and 30 lines of
Fe~II from the lists  \cite{4,5,11,12}.

{\bf Determination of absolute line shifts} involves two
problems. First, we have to reduce the FTS wavelengths of solar
lines to the absolute scale. Second, we have to choose a way of
determining FTS wavelengths from FTS observations.

The first problem was solved by referring to the absolute
wavelength scale of photographic solar spectrum derived by 
Pierce and  Breckinridge \cite{19} as well as
by using the assumption of Livingston \cite{18} that the absolute shifts
of the strong Mg I~$\lambda$~517.27~nm line in active regions
coincide with the shifts in quiet regions. The assumption is
based on the fact that the Mg I line core is formed high in the
atmosphere and its granulation shift is virtually zero, that is,
it is virtually independent of small velocity field variations
in the lower and middle photosphere. In this case the procedure 
of reducing FTS wavelengths to the absolute
solar wavelength scale  \cite{19} is quite simple. First, we
found a correction to the absolute scale \cite{19} using the line
Mg~I~$\lambda$~517.27~nm:
  $$
\Delta_{abs}=\lambda_{FTS}^{MgI}-\lambda_{P\& B}^{MgI},
 $$
where $  \lambda_{FTS}^{MgI}  $   is the FTS wavelength of the
Mg I line, $  \lambda_{P\& B}^{MgI}$  is the wavelength of the
same line in the absolute scale of  Pierce and  Breckinridge \cite{19}. 
Then we reduced  line wavelengths measured from the FTS spectra ($\lambda_{FTS}$) 
to the absolute scale:
 $$
  \lambda_{abs}^{FTS}=\lambda_{FTS}-  \Delta_{abs}.
 $$
Thus the correction of the FTS wavelengths was made for tne effects of
the relative motion of the Sun
and the Earth as well as for possible systematic scale shifts.

Now we turn to the second problem --- the determination of line
wavelengths from FTS observations,  $ \lambda_{FTS} $. Since the wavelengths $
\lambda_ {FTS} $ are referred to the absolute scale of tables
\cite{19}, they must be determined in the same way as in
\cite{19}. The wavelengths in \cite{19} were measured on
photographic plates by an operator with a special measuring
machine. After a series of numerical experiments and taking into
consideration the results of Dravins et al. \cite{12}, we
concluded that the most stable results could be obtained when $
\lambda_ {FTS} $ was found as a weighted mean wavelength of line
core (or center of gravity of line core). The line core was
defined as one-tenth of the line profile measured from the
minimum profile point. To upgrade the estimate accuracy, we
interpolated the original observations by a weighted parabola
with a 0.2~pm step.

  \begin{figure}
\centerline{\includegraphics[scale=1. ]{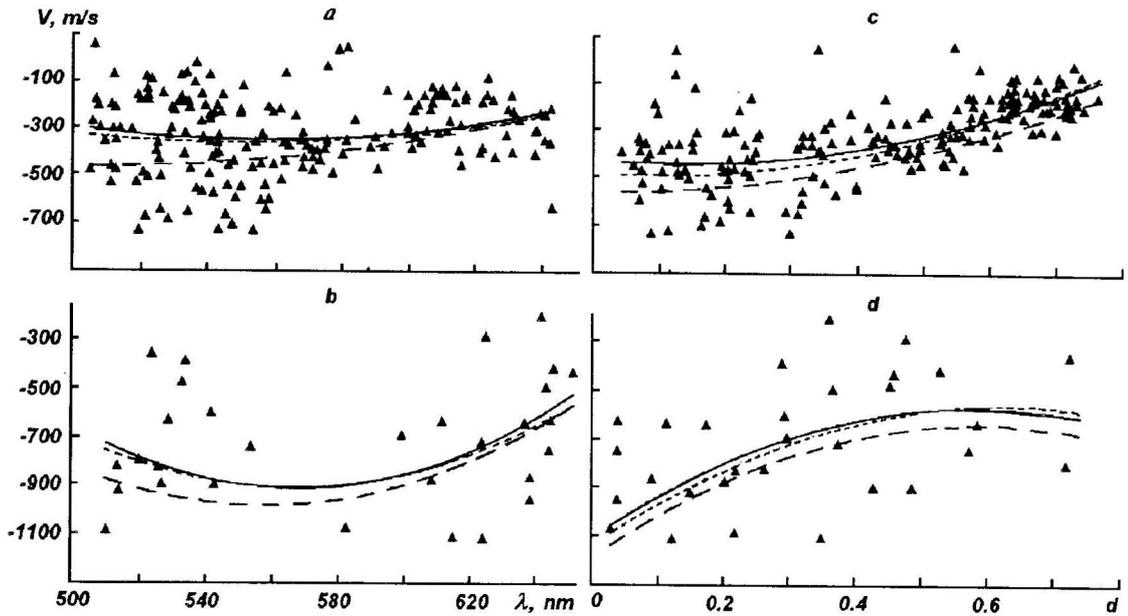}}
 \hfill
\vspace{0.1cm}
 \caption
{Absolute line shifts derived from nine FTS spectra for the
quiet photosphere in this study and from tables \cite{11,12}.
Shown is only our sample of Fe~I and Fe~II lines. Solid lines
and triangles) results from \cite{11,12}. Long dashes) shifts
with respect to the Mg~I $\lambda$~517.27~nm line derived in
this study, average over all lines; short dashes) shifts with
respect to 12 reference Fe~I lines, a and b) shifts of Fe~I and
Fe~II lines depending on wavelength, c and d) shifts depending
on central line depth.} \label{F-1}
 \end{figure}
Radial velocities were determined by the well-known expression
which relates line shifts to velocities:
 $$
V_R =c\cdot (\lambda_{abs} - \lambda_{lab}) /
\lambda_{lab}-2.12\cdot10^{-6} \cdot c,
 $$
$\lambda_{lab}$ being laboratory wavelengths \cite{11,12}. To
assess variations in the FTS dispersion, we compared the average
radial velocities derived from nine FTS spectra of quiet areas
with the data from \cite{11,12}. Recall that solar line
wavelengths in \cite{11,12} were also taken from tables \cite{19}.
The wavelength dependence of radial velocities (Fig. 1) suggests
that the FTS observations require a correction for dispersion
variations. The correction was made with the use of lines with
small granulation shifts between their positions in the active
and quiet photosphere. After preliminary calculations we
selected such lines distributed over the whole wavelength range
observed. An analysis revealed that the smallest differences in
absolute shifts between quiet and active regions occurred in two
groups of moderately strong Fe~I lines.
One group included lines with lower excitation potentials $EPL$
from 2 to 3.5~eV, equivalent widths $W$ from 7.7 to 10 pm, and
central depths $d$ from 0.65 to 0.79; the lines in the other
group have $EPL > 3.5$~eV, $W = 10.0$--12.0~pm, $d =0.70$--0.79.
Using specially calculated \cite{shem93} sensitivity indicators (coefficients
that characterize line sensitivity to variations in particular
parameters of the medium), we excluded the temperature-sensitive
lines from the list of selected lines (28 out of 189). The final
list of 12 reference Fe~I lines is given in Table 1.
%
 \begin{table}[!htb] \centering
 \parbox[b]{14cm}{
\caption{Spectral lines selected for referring the FTS
wavelength scale to the absolute scale of solar wavelengths by
Pierce and Breckinridge \cite{19}} \label{T:1} \vspace{0.3cm}}
\footnotesize
\begin{tabular}{cccccc}
 \hline
 $\lambda $, nm (laboratory) \cite{11,12}& $\lambda $, nm (solar) \cite{19}& $EPL$, eV & $d_{obs}$ &
$W_{obs}$, pm & $\log \tau_{5D}$ \\
 \hline
 509.07750  & 509.07807  & 4.26  &  0.752 &  10.04 &  -2.41\\
 514.17387  & 514.17460  & 2.42  &  0.768 &  9.80  &  -2.76\\
 519.87114  & 519.87171  & 2.22  &  0.792 &  9.96  &  -3.05\\
 525.06447  & 525.06527  & 2.20  &  0.793 & 10.35  &  -3.09\\
 536.48717  & 536.48801  & 4.44  &  0.785 & 13.18  &  -2.99\\
 538.94786  & 538.94866  & 4.41  &  0.719 &  9.57  &  -2.36\\
 546.29601  & 546.29672  & 4.47  &  0.735 &  9.70  &  -2.70\\
 548.77433  & 548.77512  & 4.14  &  0.722 &  12.02 &  -3.00\\
 557.60874  & 557.60970  & 3.43  &  0.772 &  12.30 &  -3.06\\
 566.25153  & 566.25233  & 4.18  &  0.719 &  11.16 &  -3.10\\
 624.63172  & 624.63271  & 3.60  &  0.716 &  12.37 &  -3.02\\
 641.16468  & 641.16586  & 3.65  &  0.717 &  14.32 &  -3.13\\
    \hline
\end{tabular}
\end{table}
\noindent
These lines are of low sensitivity to temperature (it is the
same for all lines --- $ (\delta d/d) / (\delta T/T) = -1.5$),
their central depths range from 0.72 to 0.79, and equivalent
widths range from 10 to 14 pm. The lines were used to find
dispersion corrections for every spectrum:
 $$
  \Delta_{dis}(\lambda)  = \lambda_{abs}^ {FeI} - \lambda_ {P  \&B}^{FeI}.
 $$
Then the dispersion curve coefficients $a$ and $b$ were found by
linear approximation:
 $$
  \Delta_{dis} (\lambda) = a + b \lambda,
 $$
and dispersion corrections were introduced to every FTS
wavelength of the other lines:
  $$
   \lambda_{abs}  =\lambda_{abs}^{FTS} - (a + b\lambda).
  $$

The radial velocities calculated with respect to 12 reference
Fe~I lines showed a much better agreement with the data by
Dravins et al. \cite{11,12} (Fig. 1).
When 12 reference Fe~I lines are used, the Mg I line can be
eliminated from the calculations. However, we determined the
radial velocities relative to the magnesium line as well, since
the change in dispersion is not very large in our case and
reference lines are not ideal for correcting it.

\section{Results}
The absolute shifts obtained by us and averaged over nine
spectra are compared in Fig. 1 with the shifts derived in
\cite{11,12}. The average cosine of the angle of radiation
direction with the external normal to the area surface, $\mu$, is
0.988 for all nine areas observed. Our results are obviously
different from those of \cite{11,12}. The discordance is
different for moderate, strong, and weak lines. We believe that
the discordance is due mainly to the procedures used for
determining line wavelengths and to the fact that the spectral
resolution was different in our observations and those in
\cite{11,12}. Nevertheless, for the radial velocities derived
with respect to the reference Fe~I lines the discordance (about
50~m/s on the average) may be reckoned as small compared to the
scatter in the data. Qualitatively the discordance is
independent of the excitation potential and ionization stage,
except for strong lines with $EPL > 3.5$~eV, most reference
lines falling in this category. The lines in this group show, of
course, better agreements with \cite{11,12}.

  \begin{figure}
\centerline{
\includegraphics [scale=1.1]{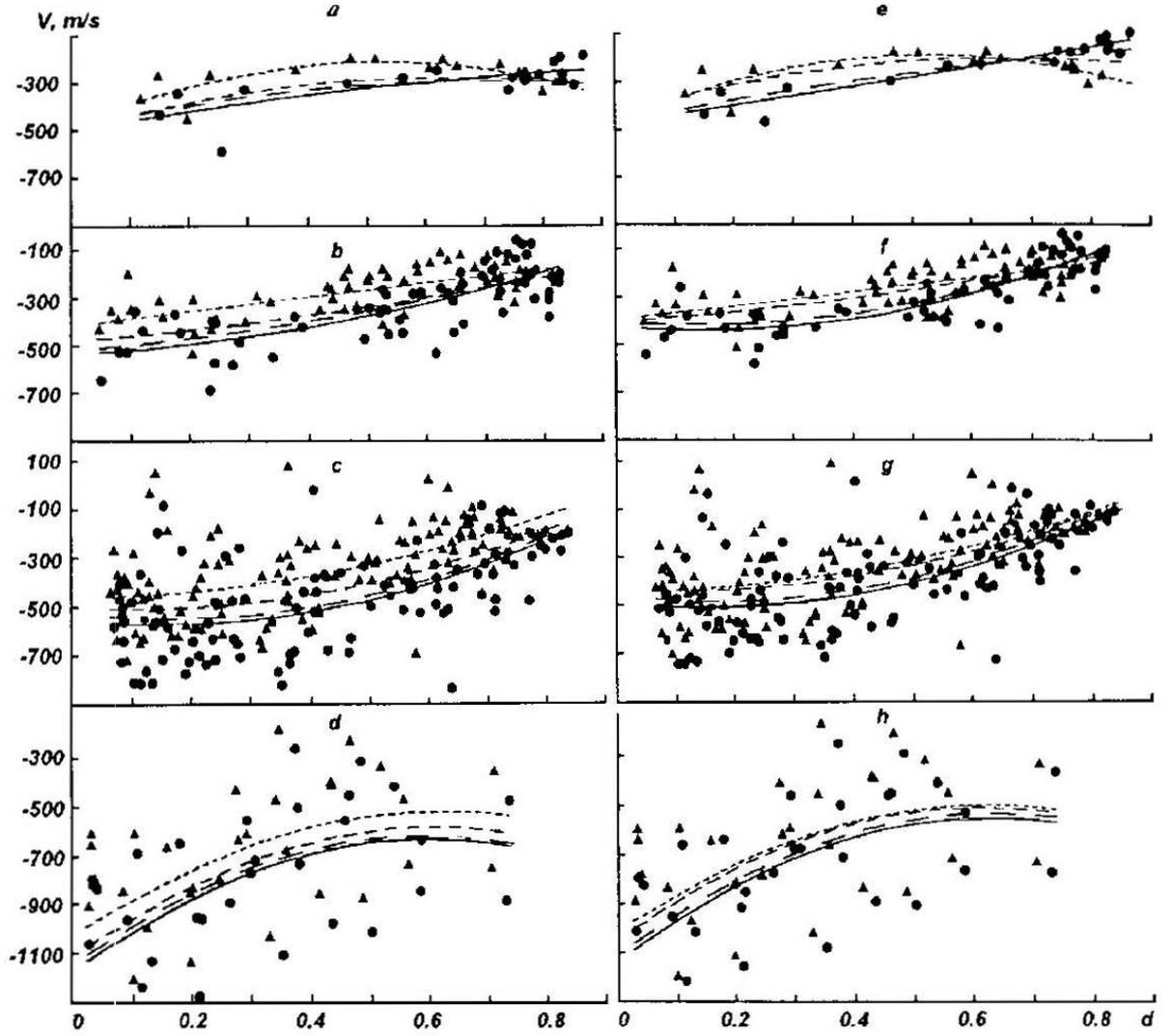}}
 \hfill
\vspace{0.1cm}
 \caption
{Average absolute line shifts with respect to the Mg I
$\lambda$~517.27~nm line (a--d) and 12 Fe~I lines (e--h) as
functions of central line depth. Central depths were determined
relative to the local continuum for each spectrum. a,e) Fe~I
lines with $EPL < 2$~eV; b,f) $EPL = 2$--3.5~eV; c,g) $EPL >
3.5$~eV; d,h) Fe~II lines. Solid lines and dots) data from nine
spectra with $\alpha \approx 0$\%, long dashes) from six spectra
with $\alpha \approx 3$~\%, short dashes) from six spectra with
$\alpha \approx 8$\%, triangles and dotted line) from two
spectra with $\alpha \approx11$\%.} \label{F-2}
 \end{figure}

Figures 2 a--d display the shifts relative to the magnesium line,
and Figs 2 e--h show the shifts relative to 12 reference Fe~I
lines. We divided 23 FTS spectra into four groups depending on
filling factor $ \alpha$: group 1 --- nine spectra,
$\overline{\alpha}=0.3$\% $\approx$~0\%, $\overline{\mu}=0.988$;
group 2 -- six spectra, $\overline{\alpha}=$2.5\%~$\approx$~3\%,
$\overline{\mu}=0.962$;  group 3 -- six spectra,
$\overline{\alpha}=7.5$\%~$\approx$~8\%, $\overline{\mu}=0.953$;
group 4 -- two spectra, $\overline{\alpha}=
11.4$\%~$\approx$~11\%, $\overline{\mu}=0.955$. The Fe~I lines
were divided into three groups in accordance with their $EPL$:
less than 2~eV, 2--3.5 eV, and more than 3.5~eV. The dependence
of absolute shifts on $\alpha$, $EPL$, line strength, and
ionization stage is apparent. First of all we note that all
Fe~II lines in active areas are red-shifted as compared to quiet
areas, the shift growing with $\alpha$. This inference is
unaffected by the choice of reference lines. The largest
relative red shifts are observed in the spectra with $\alpha
\approx 11$ per cent, they are about 100~m/s when the dispersion
is corrected and about 160 m/s when determined relative to the
magnesium line only without dispersion correction.

As to the Fe~I lines, the results heavily depend on $EPL$ and
line strength. The lines with $EPL > 3.5$~eV behave similarly to
the Fe~II lines: the shifts of most lines are more red in active
areas as compared to the quiet photosphere. This inference is
also unaffected by the choice of reference lines. At the same
time strong lines in active areas demonstrate absolute shifts
close to those observed in the quiet photosphere, this result
being the same when calculations are made with the magnesium
line or with 12 reference Fe~I lines. The tendency is stronger
for lines with smaller excitation potentials: the strong lines
either display shifts close to those in the quiet photosphere or
have more violet shifts in magnetic areas. When the change in
dispersion is allowed for, the largest violet shifts are
observed in the areas with $\alpha$ = 11 per cent, they are
about 180~m/s for strong Fe~I lines with $EPL < 2$~eV and about
80~m/s for lines with $EPL = 2$--3.5~eV. The largest red shift
is virtually the same (120~m/s) for all groups of weak and
moderate Fe~I lines.

It should be stressed that the variations of absolute shifts in
active areas with respect to their positions in quiet
photosphere areas are small and lie within confidence limits for
the corresponding mean values, and so we may speak about
tendencies only. Nevertheless, it is of interest that our
results point to the change of sign in absolute shifts of Fe~I
lines in active areas with respect to the quiet photosphere if
the line cores are formed above the level $\log
\tau_5~\approx~-3$, i.e., above 425~km. It is pertinent to note
that the absolute shifts calculated relative to the magnesium
line or strong iron lines demonstrate the same trends. This fact
substantiates the reality of the results obtained.

We have already noted that in the early study \cite{7} based
on the same FTS observations Brandt and Solanki inferred that 19
strong Fe~I lines had, on the average, larger red shifts in
active areas. The list of those lines included seven lines with
$EPL > 3.5$~eV, nine lines with $EPL = 2$--3.5~eV, and three
lines with $EPL < 2$~eV. To find out why the data from \cite{7}
disagreed with our results, we repeated the calculations made in
\cite{7}, allowing for the absolute line shifts derived in this
study. Recall that absolute shifts of bisectors of individual
lines were dealt with in \cite{7}, and then their averaging was
done. When treated in such a manner, the average bisectors were
found to have red shifts in their wings and at the intermediate
level in active areas with respect to the quiet photosphere. At
the same time, the positions of individual bisectors of the
lines studied had a strong scatter near line cores at an
intensity level of about 0.2, this scatter being due to the wide
range of $EPL$ in this line group. That is why the authors of
\cite{7} left the lower bisector part out of consideration, as
seen from Fig.~13 in \cite{7}, and formulated their conclusion
which did not characterize the behavior of line cores. We
believe, therefore, that our results are not in variance with
the results of \cite{7}, since they describe the behavior of
absolute line shifts at various levels of residual line
intensities: the shifts investigated in \cite{7} refer to the
levels from line wings to a residual intensity of 0.4, while the
shifts of line cores only are dealt with in this study.
  \begin{figure}
\centerline{
\includegraphics [scale=1.1]{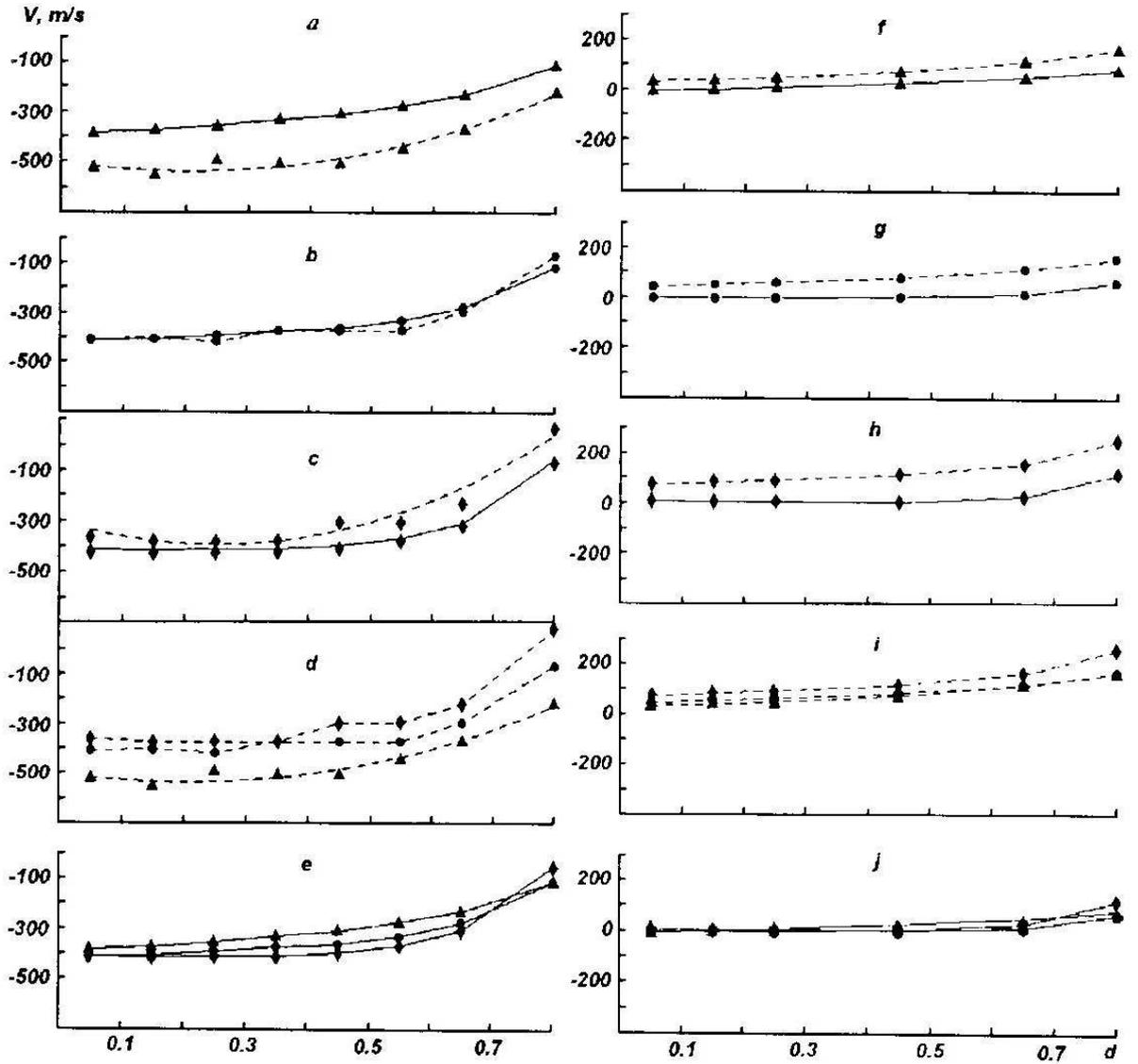}}
 \hfill
\vspace{0.1cm}
 \caption
{Absolute shifts of the artificial Fe~I lines $\lambda$~500~nm
calculated with 2-D MHD models \cite{5} (a-e) and \cite{2,3}
(f-j) as functions of central line depth. Central depths of
lines from nonmagnetic HD models are given. Solid lines) 2-D
nonmagnetic HD model calculations, dashed lines) 2-D MHD model
synthesis; triangles, a,f) lines with $EPL = 0$~eV; dots, b,g),
lines with $EPL = 2.5$~eV; diamonds, c,h) lines with $EPL =
5$~eV; d,i) lines calculated with MHD models only; e,j) lines
calculated with nonmagnetic HD models only.} \label{F-3}
 \end{figure}

As  $\overline{\mu}$ can be somewhat different for the groups of spectra, the
limb effect might show up in our results \cite{6}. To examine
this problem, we selected three spectra for areas lying at the
same distance from the disk center $\mu= 0.97$ but having
various $ \alpha$ -- 0.006, 0.026, and 0.114. The spectra were
acquired on the same day in the course of 34 min. The results
obtained from these spectra were found to be in qualitative
agreement with those obtained from 23 spectra. Thus, a possible
limb effect does not change noticeably our final results derived
from the spectra used.

\section{Interpretation of the results}

Earlier, Cavallini et al. \cite{9}  proposed their interpretation
of the line shifts in solar spectra. They suggested
that the penetrating photos pheric convection becomes less
efficient in the presence of small-scale magnetic fields. In
this case horizontal magnetic lines of force prevent ascending
flows from penetrating in the middle and upper photosphere while
the velocities grow in descending flows around magnetic tubes.
Thus, for the lines formed deep in the photosphere the violet
convective shift decreases above granules and the red shift
increases above intergranular lanes --- the line as a whole
acquires a redder absolute shift as compared to the quiet
photosphere. In the lines that form in the upper layers the
contribution to the red shift from the regions above descending
flows is smaller, since the penetrating convection is somewhat
``submerged'' in active areas.

Hanslmeier et al. \cite{13} found from spectra with high spatial
resolution that a strong line formed in a magnetic region in the
upper layers may have a redder shift above a granule than above
intergranular lanes. In other words, inversion may occur in the
motions in the upper photosphere. At the same time, a line
formed in deeper layers in the same area displays the usual
pattern: its shift is violet above granules and red above
porules.

Our quantitative interpretation is based on two sets of
theoretical self-consistent two-dimensional MHD models. The
models and some results of their application to spectral
observations are described in \cite{2,3,5}. Three sets of models
with mean magnetic fluxes of 10, 20, and 30 mT were presented in
papers \cite{2,3}.
%
 \begin{table}[!htb] \centering
 \parbox[b]{14cm}{
\caption{Basic parameters of the 2-D MHD models used
 \label{T:2} }
\vspace{0.3cm}} \footnotesize
\begin{tabular}{ccc}
 \hline
  Model parameter  & Atroshchenko \&  Sheminova \cite{5}  &   Brandt \& Gadun \cite{2} \\
 \hline
 Area size ($x,~z$),~km  &    2520$\times$2030&      3840$\times$1920\\
 Grid dimension, Step on $x,~z$ & 72$\times$58, 35~km& $256\times128$, 15~km\\
 Model atmosphere layers (above  $\tau$=1)&   900~km&  600~km\\
 Radiation transfer&     2D momentum method,& 2D momentum method, \\
                   & variable Eddington factors&variable Eddington   factors\\
 Absorption coefficient &  monochromatic, 97 frequencies,& Rosseland coefficient,  \\
 in photospheric layers &absorption taken into &selective absorption corrected  \\
                      &  account directly (ODF)& in four frequency intervals\\
 Side boundary conditions&    symmetric impenetrability&    periodic\\
 Upper boundary conditions&   free&    free\\
 Lower boundary conditions&   free&    closed\\
 Mean magnetic field strength in the model&       45 mT&  30 mT\\
 Magnetic field strength maximum &    300 mT&  200 mT\\
at the level $\tau$~=~1 in a  magnetic tube&&\\
 Largest Wilson depression in a tube&   $\approx$300~km&    $\approx$200~km\\
 Magnetic tube width at $\tau$~=~1 &     $\approx$300~km&  $\approx$250~km\\
  \hline
     \end{tabular}
\end{table}
\noindent
We used the model with the largest flux. Table 2 gives the most
important model parameters. 

We calculated the sets of artificial
Fe~I and Fe~II lines with the wavelength $  \lambda$~500~nm and
excitation potentials of 0, 2.5, 5~eV (Fe~I) and 2, 3.5, 6~eV
(Fe~II). The LTE line profiles of these lines were calculated
with the SPANSAT code \cite{spansat}. First we took the
corresponding two-dimensional inhomogeneous hydrodynamic models
calculated with the same approximations as in MHD models but
without magnetic fields and found such values of $Agf$ (the
product of abundance by oscillator strength) that fixed central
line depths might be obtained: 0.05, 0.15, ..., 0.65, 0.8 for
Fe~I and 0.05, 0.15, ..., 0.65 for Fe~II. Then these values of
$Agf$ were used to calculate lines with 2-D MHD models in the
LTE approximation without broadening by magnetic fields. Figures
3 and 4 show the absolute shifts thus obtained for Fe~I and
Fe~II lines, respectively. 

First of all we call attention to the
fact that models \cite{2,3} have overheated atmospheres because
the absorption by spectral lines was overestimated in them, and
this results in a smaller vertical velocity gradient. That is
why the lines calculated with 2-D HD models \cite{2,3} have
underestimated violet shifts.

  \begin{figure}
\centerline{
\includegraphics [scale=1.]{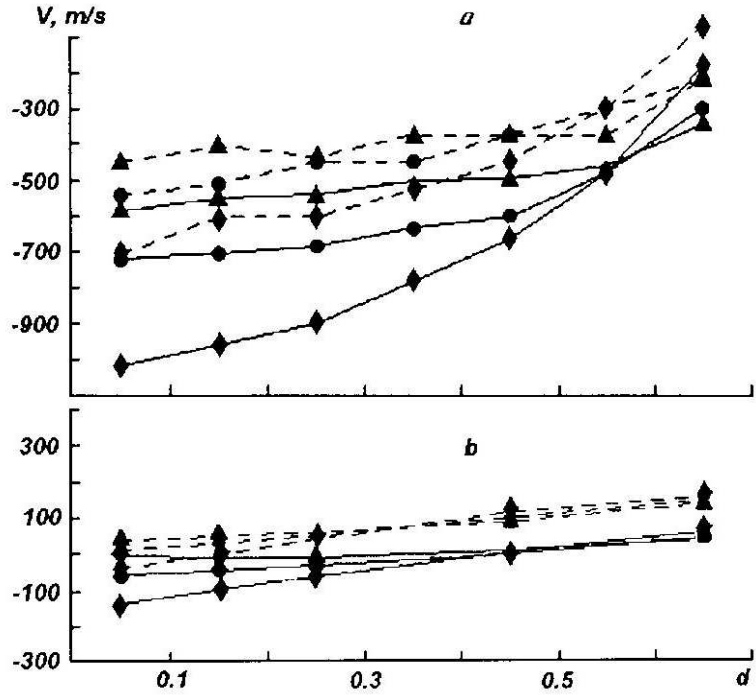}}
 \hfill
\vspace{0.1cm}
 \caption
{Absolute shifts of the artificial Fe~II lines  $\lambda$~500~nm
calculated with 2-D MHD models \cite{5} (a) and \cite{2,3} (b).
Solid lines) calculations with 2-D nonmagnetic HD models, dashed
lines) 2-D MHD model synthesis, triangles) lines with $EPL =
2$~eV, dots) $EPL = 3.5$~eV, diamonds) $EPL = 6$~eV.}
\label{F-4}
 \end{figure}
The absolute line shifts are qualitatively the same in both sets
of nonmagnetic HD models (Figs 3e and 3j). The shifts of
moderate-strong lines depend on $EPL$ -- the violet shift is
larger the higher the excitation potential (i.e., the relative
contribution from ascending flows is greater for deep-formed
lines). The models for strong Fe~I lines reproduce the
well-known effect --- lines with high $EPL$ display redder shifts
due to an inverse distribution of temperature variations in the
middle and upper photosphere. These patterns are in good
qualitative agreement with observations \cite{12}.

The 2-D MHD models demonstrate the reverse dependence of line
shifts on $EPL$ (Figs 3d and 3i: the shifts become redder with
growing $EPL$, reflecting the growing role of descending flows
in the lower layers. However, the shifts of Fe~I lines from MHD
models relative to the corresponding nonmagnetic HD models (this
corresponds to an observational comparison between active areas
and quiet photosphere areas) are represented in different ways
by two sets of MHD models. Models \cite{2,3} show a slightly
growing relative red shift (up to 100~m/s) between HD and MHD
models when $EPL$ increases. The other set of MHD models
\cite{5} show a more intricate pattern: the lines with $EPL = 0$
are violet-shifted, the shift is almost the same for all lines;
the shifts of the Fe~I lines with $EPL = 2.5$~eV found from MHD
models equal the shifts from nonmagnetic HD models; and the
lines with high excitation potentials (5~eV) have red shifts.

There are no substantial discrepancies between the models for
Fe~II lines (Fig. 4). In both model sets the MHD model lines are
red-shifted with respect to nonmagnetic HD models.

So, we may state that two-dimensional MHD models \cite{2,3,5}
reproduce qualitatively alike the red shifts of the lines formed
deep in the photosphere. The shifts are due to growing areas of
descending flows, where bundles of magnetic lines of force are
located, and to growing velocities of descending flows. However,
there is essential discordance between the models when lines
formed in the middle and upper photosphere are considered. The
discordance owes its existence to some drawbacks of both model
sets. Models \cite{5} have a very rough spatial step -- 35~km,
and this could distort the temperature distribution in the
middle and upper photosphere when a developed magnetic
configuration was calculated. Models \cite{2,3} overestimate the
heating of atmospheric layers due to an overestimated absorption
in line frequencies, and this also may distort the velocity
field. Thus, the violet shifts found by Cavallini et al.
\cite{9} and in our study for the lines that form in the upper
layers are not confirmed within the scope of the 2-D MHD models
used by us, although these models need serious improvements.

\section{Conclusion}

FTS observations were used to estimate possible absolute shifts
of Fe~I and Fe~II lines in active regions. We conclude that weak
Fe~I lines and all Fe~II lines are likely to be red-shifted with
respect to their positions in the quiet photosphere, while
strong Fe~I lines with low $EPL$s, their cores being formed
above the level $\log \tau_5\approx-3$ ($\approx$425~km), have
more violet shifts.

The results were interpreted on the basis of 2-D MHD models that
reproduce quite adequately the red shifts of the lines formed
deep in the photosphere, but violet shifts of the lines formed
in the upper photospheric layers are not confirmed within the
scope of these models.

\vspace{1.cm}
 {\bf Acknowledgements.}
We wish to thank the National Optical Astronomical
Observatory/National Solar Observatory (Tucson, Arizona, USA)
and especially J. Brault and B.~Graves for support in the FTS
observations; S. Solanki for filling factor calculations; R.~Rutten 
for assistance in the organization of this investigation.
The study was partially financed by the Joint Foundation of the
Government of Ukraine and International Science Foundation
(Grant No. K11100).



\end{document}